\documentclass[11pt]{article}
\usepackage{fullpage}
\usepackage[dvips]{epsfig}

\def\##1{{\bf #1}}
\def\=#1{\underline{\underline{#1}}}

\def\+#1{\underline{\bf #1}}
\def\*#1{\underline{\underline{\bf #1}}}

\def\r#1{(\ref{#1})}
\def\l#1{\label{#1}}
\def\c#1{\cite{#1}}

\def\le{\left(}
\def\ri{\right)}
\def\les{\left[}
\def\ris{\right]}
\def\lec{\left\{}
\def\ric{\right\}}

\def\.{\mbox{ \tiny{$^\bullet$} }}

\def\epso{\epsilon_{\scriptscriptstyle 0}}

\def\muo{\mu_{\scriptscriptstyle 0}}

\def\O{\scriptscriptstyle 0}

\def\eps{\epsilon}

\begin{document}

\begin{center}
\Large{\bf {\Large  
Geometrically-Derived Anisotropy in\\
Cubically Nonlinear Dielectric Composites
}}

\normalsize
\vspace{6mm}

Tom G. Mackay\footnote{ Tel: +44 131 650 5058; fax: +44 131 650 6553;
e--mail: T.Mackay@ed.ac.uk}

\vspace{4mm}

\noindent{ \emph{School of Mathematics, The University of
Edinburgh, James Clerk Maxwell Building, \\The King's Buildings,
Edinburgh EH9 3JZ, United Kingdom.} }

\vspace{28mm}

{\bf Abstract}\end{center}

We consider an anisotropic homogenized composite medium (HCM)
 arising from isotropic particulate 
component phases
based on ellipsoidal geometries. 
For cubically nonlinear component
phases, the corresponding zeroth-order strong-permittivity-fluctuation
theory (SPFT) (which is equivalent to
the Bruggeman homogenization formalism) and second-order 
SPFT are established and used to estimate
the constitutive properties of the HCM. The relationship between the
component phase particulate geometry and the HCM constitutive
properties is explored. Significant differences are highlighted
between the estimates of the Bruggeman homogenization formalism and
the second-order SPFT estimates. The prospects for nonlinearity
enhancement
 are investigated.

\vspace{24mm}

\section{Introduction}

The constitutive properties of a homogenized composite medium (HCM) are
determined by both the constitutive properties and the topological
properties of its component phases \c{L96}--\c{M03}. In particular,
component phases based on nonspherical particulate geometries may give
rise to
 anisotropic HCMs, despite the  component phases  themselves being
 isotropic with respect to
their electromagnetic properties. Such 
geometrically-derived anisotropy has been extensively characterized
for linear dielectric  HCMs \c{Ward}--\c{MW_biax2} and more general
bianisotropic HCMs \c{MW_biax2}--\c{MW02}.
For weakly nonlinear HCMs, the role of the component phase particulate
geometry
was   emphasized recently in this journal by Goncharenko,
Popelnukh and Venger \c{GPV},
using an approach founded on 
the mean-field approximation. However, their analysis
was restricted to the Maxwell Garnett homogenization formalism
\c{Boyd96, Zeng}. 
A more comprehensive study is communicated here based on the
strong-permittivity-fluctuation theory (SPFT) \c{TK81}. 
In contrast to the
aforementioned  Maxwell Garnett approach \c{GPV}, the SPFT approach (i)
incorporates higher-order  statistics to describe the component
phase distributions; (ii) is not restricted to only dilute composites; and (iii)
is not restricted to only weakly nonspherical
particulate geometries.

The early development of the  SPFT
concerned wave propagation in continuous random mediums \c{Ryzhov, Frisch}, 
but more recently the theory has been applied to the estimation of 
HCM constitutive
parameters \c{L01, MLW02a}. 
The  SPFT represents a significant advance over conventional homogenization
formalisms, such as the Maxwell Garnett approach and the Bruggeman
approach \c{Boyd96, Ward},
through incorporating a comprehensive description of the
distributional statistics of the HCM component phases.
In estimating the constitutive parameters of an HCM,
the SPFT
employs
 a Feynman--diagrammatic technique to
 calculate iterative refinements to the constitutive parameters of a 
comparison medium; successive
iterates incorporate successively higher--order 
spatial correlation functions. 
It transpires that the SPFT 
comparison medium is equivalent to the effective medium
of the (symmetric) Bruggeman homogenization theory \c{ML95, MLW00}.
In principle, correlation functions of
arbitrarily high order can be accommodated in the SPFT. However,
the theory is most widely-implemented at the level of the \emph{bilocal
approximation} (i.e., second-order approximation), wherein a two-point covariance
function and its associated correlation length $L$ characterize the
component phase distributions. As indicated in figure~1, 
coherent interactions between pairs of
scattering centres
within a region of linear dimensions
$L$ are incorporated in the bilocal SPFT; 
scattering centres separated by distances much greater than
$L$ are assumed to act independently. Thereby, the SPFT provides an
estimation of coherent scattering losses, unlike the Maxwell Garnett
and Bruggeman homogenization formalisms. Notice that
 the bilocally--approximated SPFT gives
rise to the Bruggeman homogenization formalism in the limit 
$L \rightarrow 0$ \c{MLW00}.

The SPFT has been widely applied to linear homogenization scenarios, where
generalizations\footnote{The generalized SPFT is referred to as the
\emph{strong-property-fluctuation theory}.}
have been  developed for anisotropic dielectric \c{Zhuck, MLW01b}, isotropic
chiral \c{ML95} and bianisotropic \c{MLW00, MLW01a} HCMs. 
Investigations of the trilocally-approximated
SPFT for isotropic HCMs have recently confirmed the convergence 
of the second-order theory \c{MLW02a, MLW01c, MLW02b}.
In the weakly nonlinear regime,   developments of the
bilocally-approximated SPFT have  been  restricted to 
  isotropic HCMs, based on spherical component phase geometry 
  \c{L01, MLW02a, MLW02b}.
The present study  advances the nonlinear SPFT through developing
the theory for  cubically nonlinear, anisotropic 
HCMs. Furthermore,  
it is assumed that the 
component phases are composed of
electrically-small ellipsoidal
particles. 
The relationship between the HCM constitutive parameters
and the underlying particulate geometry of the 
component phases  is investigated
via a representative numerical example.

In our notational convention, dyadics are double underlined
whereas vectors are in bold face.
 The inverse, adjoint, determinant and trace of
a dyadic $\=A$ are denoted by $\=A^{-1}$, $\=A^{adj}$ ,
$\mbox{det}\,\les \, \=A \, \ris $ and $\mbox{tr}\,\les \, \=A \,  \ris $, respectively.
The  identity dyadic
 is represented by  $\,\=I\,$.
 The ensemble average of a quantity $\psi$ is
written as $\langle \, \psi \, \rangle$.
 The permittivity and permeability of free
space (i.e., vacuum) are given
by $\epso$ and $\muo$, respectively; $k_{\O} = \omega \sqrt{\epso \muo}$
is the free-space wavenumber while  $\omega$ is the  angular frequency.

\section{Homogenization generalities}

\subsection{Component phases}

Consider the homogenization of a two-phase composite with
 component phases  labelled as  $a$ and $b$. The component
phases are taken  to be isotropic dielectric mediums with 
permittivities
\begin{equation}
\eps_\ell  = \eps_{\ell\,0} + \chi_\ell \, |\, \#E_{\,\ell} \,|^2,
\hspace{30mm} (\ell = a,b), \l{comp_const}
\end{equation}
where $\eps_{\ell\,0}$ is the  linear permittivity,
$\chi_\ell$ is the nonlinear susceptibility, and $|\, \#E_{\,\ell}
\,|^2$ is the electric field developed inside a region of phase
$\ell$ by illumination of the composite medium. We assume weak
nonlinearity; i.e., $|\,\eps_{\ell\,0}\,| \gg |\, \chi_\ell\,| \,
|\, \#E_{\, \ell} \,|^2$. Notice that such electrostrictive mediums as
characterized by \r{comp_const} can induce Brillouin scattering which
is often a strong process \c{Boyd}.
The component phases $a$ and $b$ are taken to be randomly distributed as
identically-orientated, conformal ellipsoids. The shape dyadic 
\begin{equation}
\=U = 
\frac{1}{\sqrt[3]{U_x U_y U_z}}
\;
\mbox{diag} (U_x,
U_y, U_z),\hspace{25mm} (U_x, U_y, U_z > 0), \l{comp_U}
\end{equation}
parameterizes the conformal ellipsoidal surfaces as
\begin{equation}
\#r_{\,e}(\theta,\phi) = \eta \, \=U\.\#{\hat r} \, (\theta,
\phi),
\end{equation}
where  $\#{\hat r} \, (\theta, \phi)$ is the radial unit vector
specified by the spherical polar coordinates $\theta$ and $\phi$.
Thus, a wide range of
ellipsoidal particulate shapes, including highly elongated forms, can
be accommodated.
 The linear ellipsoidal dimensions, as determined by $\eta$, are
assumed to be 
sufficiently small that
the electromagnetic long-wavelength regime pertains.

In the SPFT, statistical moments of the characteristic functions
\begin{equation}
\Phi_{ \ell}(\#r) = \left\{ \begin{array}{ll} 1, & \qquad \#r \in
V_{\, \ell},\\ & \qquad \qquad \qquad \qquad \qquad \qquad (\ell=a,b) , \\
 0, & \qquad \#r \not\in V_{\, \ell}, \end{array} \right.
\end{equation}
are utilized to
 take account of the
 component phase distributions. 
 The volume fraction of phase $\ell$, namely $f_\ell$ , is given by
the first statistical moment of
 $\Phi_{\ell}$ ;
 i.e., $\langle \, \Phi_{\ell}(\#r) \, \rangle = f_\ell$ .
 Clearly,
 $f_a + f_b = 1$. The second statistical moment of $\Phi_{\ell}$
 provides
a two-point covariance function; we adopt the
physically-motivated form \c{TKN82}
\begin{equation}
\langle \, \Phi_\ell (\#r) \, \Phi_\ell (\#r')\,\rangle = f_\ell
\, \les 1 + \le \, f_\ell - 1\,\ri\, \mathcal{H} \le \,\sigma -
L\,\ri\,\ris\,, \l{cov}
\end{equation}
where $\mathcal{H}$ is the Heaviside function (i.e., $\mathcal{H}
(x) = \int^x_{-\infty} \delta(y)\,dy$ where $\delta$ is the Dirac
delta function), $\sigma = |\,\=U^{-1}\.\#R\,|$ with $\#R = \#r - \#r'$, and
$L>0$ is the correlation length. 
The specific nature of the covariance function has been found to exert
little influence on the SPFT estimates for  linear \c{MLW01b}
and weakly nonlinear \c{MLW02a} HCMs.

\subsection{Homogenized composite medium}

Let $\#E_{\,HCM}$ denote the spatially-averaged electric field in
the HCM.
In this communication we derive the estimate 
\begin{eqnarray}
\=\eps_{\,ba}  &=& \=\eps_{\,ba0} + \=\chi_{\,ba} \, |\,
\#E_{\,HCM} \,|^2  \l{ba1} \\
&=& \mbox{diag} \le \eps^x_{ba0}, \,\eps^y_{ba0}, \,\eps^z_{ba0}
\ri +
\mbox{diag} \le
 \chi^x_{ba},\, \chi^y_{ba},\, 
\chi^z_{ba} \ri \, |\, \#E_{\,HCM}
\,|^2 \l{ba2}
\end{eqnarray}
of the HCM permittivity. The  
 \emph{bilocally-approximated}
 SPFT is utilized (hence the subscripts \emph{ba} in \r{ba1}, \r{ba2} ). 
Note that
the Bruggeman estimate of the HCM permittivity, namely
\begin{eqnarray}
\=\eps_{\,Br}  &=& \=\eps_{\,Br0} + \=\chi_{\,Br} \, |\,
\#E_{\,HCM}
\,|^2\\
&=& \mbox{diag} \le \eps^x_{Br0}, \,\eps^y_{Br0}, \,\eps^z_{Br0}
\ri +
\mbox{diag} \le
 \chi^x_{Br},\, \chi^y_{Br},\, 
\chi^z_{Br} \ri \, |\, \#E_{\,HCM}
\,|^2,
\end{eqnarray}
characterizes the comparison
medium which is adopted in the bilocally-approximated SPFT \c{MLW00}.
As the Bruggeman homogenization formalism~---~in which the component
phases $a$ and $b$ are treated symmetrically \c{Ward}~---~provides the 
comparison medium, the SPFT homogenization 
approach (like the Bruggeman formalism)  is
applicable for all volume fractions $f_a \in (0,1)$.

\subsection{Depolarization and polarizability dyadics}

The depolarization dyadic $\=D$ is a key element in both  Bruggeman and
SPFT homogenizations. It provides the electromagnetic response of a
$\=U$-shaped exclusion volume, immersed in
a homogeneous background, in the limit $\eta \rightarrow 0$.
For the component phases described by \r{comp_const} and
\r{comp_U}, we find \c{M97, MW97}
\begin{eqnarray}
\=D &=& \frac{1}{ i\, \omega \, 4 \pi} \, \int^{2\pi}_0 \; d\phi
\, \int^\pi_0\;  d\theta\; \sin \theta \, \le \,
\frac{1}{\mbox{tr} \les \,\=\eps_{\,Br}\.\=A \,\ris} \, \=A \,
\ri\,, \l{depol}
\end{eqnarray}
 wherein
\begin{equation}
\=A = \mbox{diag} \,\le \, \frac{\sin^2 \theta \,
\cos^2\phi}{U^2_x},\, \frac{\sin^2 \theta \,
\sin^2\phi}{U^2_y},\,\frac{\cos^2\theta}{U^2_z}\,\ri\,.
\end{equation}
The integrations of \r{depol}
 reduce to elliptic function representations \c{W98}. In the
case of spheroidal particulate geometries, hyperbolic 
 functions
provide an evaluation of $\=D$ \c{M97}, while for the degenerate
isotropic case  $U_x = U_y = U_z$ we have the well-known
result $\=D = \le 1 / i \omega 3 \ri \=\eps^{-1}_{\,Br}$ \c{BH}.
 We express $\=D$ as the  sum of linear and weakly
nonlinear parts
\begin{equation}
\=D = \=D_{\,0} + \=D_{\,1} \, | \, \#E_{\,HCM} \, |^2,
\end{equation}
with
\begin{eqnarray}
&& \=D_{\,0} = \frac{1}{ i\, \omega \, 4 \pi} \, \int^{2\pi}_0 \;
d\phi \, \int^\pi_0\;  d\theta\; \sin \theta \, \le \,
\frac{1}{\mbox{tr} \les \,\=\eps_{\,Br0}\.\=A \,\ris} \, \=A \,
\ri\,,
 \\&&
 \=D_{\,1} = -
\frac{1}{ i\, \omega \, 4 \pi} \, \int^{2\pi}_0 \; d\phi \,
\int^\pi_0\;  d\theta\; \sin \theta \, \les \, \frac{\mbox{tr}
\les \,\=\chi_{\,Br}\.\=A \,\ris}{\le \,\mbox{tr} \les
\,\=\eps_{\,Br0}\.\=A \,\ris \,\ri^2} \, \=A \, \ris\,.
\end{eqnarray}

A convenient construction in homogenization formalisms
 is the polarizability dyadic $\=X_{\,\ell}$, defined as
\begin{equation}
\=X_{\,\ell} = -i\,\omega\,\le\,\eps_{\ell}\,\=I -
\=\eps_{\,Br}\,\ri\.\=\Gamma^{-1}_{\,\ell}\,, \hspace{30mm} (\ell
=a,b), \l{X_def}
 \end{equation}
where
\begin{equation}
\=\Gamma_{\,\ell} =  \les\, \=I + i \omega \, \=D\.\le\,
\eps_{\ell}\,\=I - \=\eps_{\,Br}\,\ri\,\ris\,. \l{e16}
\end{equation}
Let us proceed to calculate the linear and nonlinear 
contributions  in the decomposition 
\begin{eqnarray}
\=X_{\,\ell} &=& \=X_{\,\ell \, 0} +  \=X_{\,\ell
\, 1}\, | \, \#E_{\,HCM} \, |^2\,, \hspace{30mm} (\ell = a,b).
\end{eqnarray}
Under the assumption of weak nonlinearity, we express \r{e16} in the form
\begin{eqnarray}
\=\Gamma_{\,\ell} &=& \=\Gamma_{\,\ell \, 0} +  \=\Gamma_{\,\ell
\, 1}\, | \, \#E_{\,HCM} \, |^2\,, \l{Gamma}
\end{eqnarray}
 with linear term
\begin{eqnarray}
&&\=\Gamma_{\,\ell\,0} =
 \mbox{diag} \le \Gamma^x_{\ell\, 0}, \,\Gamma^y_{\ell\,
0},\,\Gamma^z_{\ell \, 0} \ri =
  \=I + i \omega \, \=D_{\,0}\.\le\,
\eps_{\ell\,0}\,\=I - \=\eps_{\,Br0}\,\ri\,,
\end{eqnarray}
and nonlinear term
\begin{eqnarray}
&& \=\Gamma_{\,\ell\,1} =
\mbox{diag}
 \le \Gamma^x_{\ell \, 1}, \,\Gamma^y_{\ell \, 1},\,\Gamma^z_{\ell \, 1} \ri =
 i \omega \,\les \, \=D_{\,0}\.\le\,
g_\ell \, \chi_{\ell}\,\=I - \=\chi_{\,Br}\,\ri + \=D_{\,1}\.\le\,
\eps_{\ell\,0}\,\=I - \=\eps_{\,Br0}\,\ri \,\ris\,. \l{Gamma1}
\end{eqnarray}
The local field factor
\begin{equation}
g_\ell = \frac{d \, |\,\#E_{\ell} \, |^2}{ d \, |\,\#E_{\,HCM} \,
|^2}\,, \hspace{30mm} (\ell = a,b),
\end{equation}
has been incorporated in deriving \r{Gamma}--\r{Gamma1}, via the
Maclaurin series expansion $ \eps_\ell = \eps_{\ell\,0} + g_\ell\,\chi_\ell \,
|\, \#E_{\,HCM}\,|^2$. An appropriate  estimation of the local field
factor is provided by
\c{LL01}
\begin{equation}
g_\ell =  \left| \,
 \frac{1}{3}\, \le \,
 \mbox{tr} \les \,
\=\Gamma^{-1}_{\,\ell\, 0} \,\ris\,\ri \,
 \right|^2.
\end{equation}
Thus, the inverse of $\=\Gamma_{\,\ell}$ is given as 
\begin{eqnarray}
\=\Gamma^{-1}_{\,\ell} &=& \=\Gamma^{-1}_{\,\ell \, 0} +
\=\Lambda_{\,\ell }\, | \, \#E_{\,HCM} \, |^2\,, \l{e22}
\end{eqnarray}
wherein
\begin{eqnarray}
&&
\=\Lambda_{\,\ell} = \frac{1}{\det \les \,
\=\Gamma_{\,\ell \, 0}\,\ris}\les \, \mbox{diag} \, \Big( \,
 \Gamma^y_{\ell \, 1} \Gamma^z_{\ell \, 0} + \Gamma^y_{\ell \, 0} \Gamma^z_{\ell \, 1},\,
 \Gamma^z_{\ell \, 1} \Gamma^x_{\ell \, 0} + \Gamma^z_{\ell \, 0} \Gamma^x_{\ell \, 1},\,
 \Gamma^y_{\ell \, 1} \Gamma^x_{\ell \, 0} + \Gamma^y_{\ell \, 0} \Gamma^x_{\ell \, 1}
\,\Big) - \rho_\ell \, \=\Gamma^{-1}_{\,\ell\,0} \, \ris,\nonumber \\
&& \l{e23}
\end{eqnarray}
and
\begin{eqnarray}
\rho_\ell = \Gamma^x_{\ell \, 0} \Gamma^y_{\ell \, 0}  \Gamma^z_{\ell
\, 1} + \Gamma^x_{\ell \, 0}
 \Gamma^y_{\ell \, 1}
\Gamma^z_{\ell \, 0} + \Gamma^x_{\ell \, 1} \Gamma^y_{\ell \, 0}  \Gamma^z_{\ell \, 0}\,.
\end{eqnarray}
Combining \r{e22} and \r{e23} with \r{X_def}, and separating
linear and nonlinear terms, provides
\begin{equation}
\left.
\begin{array}{l}
\=X_{\,\ell \,0} = -i\,\omega\,\le\,\eps_{\ell\, 0}\,\=I -
\=\eps_{\,Br0}\,\ri\.\=\Gamma^{-1}_{\,\ell \,0}
\\
\=X_{\,\ell \,1} = -i\,\omega\,
\les \,
\le\,g_\ell \, \chi_{\ell}\,\=I -
\=\chi_{\,Br}\,\ri\.\=\Gamma^{-1}_{\,\ell \,0}\, +
\le\,\eps_{\ell\, 0}\,\=I -
\=\eps_{\,Br0}\,\ri\.\=\Lambda_{\,\ell }\,\ris\,
\end{array}
\ric
, \hspace{15mm} (\ell = a,b).
\end{equation}

\subsection{Bruggeman homogenization}

The Bruggeman estimates  of the HCM  linear permittivity  $\=\eps_{\,Br
0}$ and nonlinear susceptibility $\=\chi_{\,Br}$
are  delivered through   solving the nonlinear equations \c{Boyd96, Ward, LL01}
\begin{equation}
 f_a \, \=X_{\,a \mbox{j}} + f_b \, \=X_{\,b \mbox{j}} = \=0\,,
 \hspace{30mm} (\,\mbox{j}=0,1).
 \end{equation}
Recursive procedures for this purpose provide
the $p^{\mbox{th}}$ iterates \c{M00, MLW02b}
\begin{equation}
\left.\begin{array}{l}
\=\eps_{\,ba0} \,[p] = \mathcal{T}_{\eps}
 \, \lec \,  \=\eps_{\,ba0} \,[p-1] \, \ric\\
\=\chi_{\,ba} \,[p] = \mathcal{T}_{\chi}
 \, \lec \,  \=\chi_{\,ba} \,[p-1] \, \ric
\end{array} \ric
\end{equation}
in terms of the $(p-1)^{\mbox{th}}$ iterates, wherein
the operators $\mathcal{T}_{\eps ,\chi}$ are defined by 
\begin{eqnarray}
\mathcal{T}_\eps \, \lec \, \=\eps_{\,ba0} \,  \ric 
&=& \le \, f_a \, \eps_{a0}
\, \=\Gamma^{-1}_{\,a0} +  f_b \,  \eps_{b0} \, \=\Gamma^{-1}_{\,b0} \, \ri
\.
 \le \, f_a \,  \=\Gamma^{-1}_{\,a0} +  f_b \, \=\Gamma^{-1}_{\,b0} \, \ri^{-1}\,,
\l{tau0} \nonumber \\
 \mathcal{T}_\chi \, \lec \, \=\chi_{\,ba} \,  \ric &=&
\lec
 f_a \les g_a\,\chi_{a} \, \=\Gamma^{-1}_{\,a0} + \le \eps_{a0} \,
\=I -
\=\eps_{\,ba0} \ri \. \=\Lambda_{\,a}
 \ris   +
 f_b \les g_b\,\chi_{b} \, \=\Gamma^{-1}_{\,b0} + \le \eps_{b0} \,
\=I -
\=\eps_{\,ba0} \ri \. \=\Lambda_{\,b}
 \ris \ric
\nonumber
 \\
& &  
 \. \le f_a \, \=\Gamma^{-1}_{\,a0} + f_b \, \=\Gamma^{-1}_{\, b0} \ri^{-1} \, ,
\l{tau1} 
\end{eqnarray}
while suitable initial values are given by
\begin{equation}
\left.\begin{array}{l}
\=\eps_{\,ba0} \,[0] = \le \, f_a \,\eps_{a0} +
f_b \, \eps_{b0} \, \ri \, \=I \\
  \=\chi_{\,ba} \,[0] = \le \, f_a \, \chi_{a} + f_b \, \chi_{b}
\, \ri \, \=I
\end{array}\ric\,.
\l{init_val}
\end{equation}

\section{The bilocally-approximated SPFT}

The bilocally-approximated SPFT estimate of the HCM  permittivity
dyadic, as derived elsewhere  \c{MLW00},  is
given by
\begin{equation}
\=\eps_{\,ba} = \=\eps_{\,Br} - \frac{1}{i \,\omega}\,\le\,\=I +
\=\Sigma_{\,ba} \. \=D \,\ri^{-1}\.\=\Sigma_{\,ba}\,;
\end{equation}
the \emph{mass operator} term
\begin{equation}
\=\Sigma_{\,ba} = \le\,\=X_{\,a} -
\=X_{\,b}\,\ri\.\=W\.\le\,\=X_{\,a} - \=X_{\,b}\,\ri 
\end{equation}
is specified in terms of the  principal value integral
\begin{equation}
\=W = \mathcal{P} \,\int_{\sigma \leq L} \; d^3 \#R \: \;
\=G_{\,Br}(\#R ), \l{W_def}
\end{equation}
with  $\=G_{\,Br} (\#R)$ being
 the unbounded dyadic Green
function of the comparison medium.
Here  we develop expressions for the linear and nonlinear 
contributions of $\=\eps_{\,ba}$, appropriate to the component phases
specified in \S2.

Under the assumption of weak nonlinearity, we express 
$\=W = \=W_{\,0} + \=W_{\,1}\, | \, \#E_{\,HCM} \, |^2$; 
integral expressions for $\=W_{\,0}$ and $\=W_{\,1}$ are provided in
the Appendix. Thereby, 
the linear and nonlinear terms in the mass operator  decomposition
$\=\Sigma_{\,ba}
=  \=\Sigma_{\,ba0} + \=\Sigma_{\,ba1}\, | \, \#E_{\,HCM} \, |^2$
are given as
\begin{eqnarray}
&& \=\Sigma_{\,ba0} = \le\,\=X_{\,a0} -
\=X_{\,b0}\,\ri\.\=W_{\,0}\.\le\,\=X_{\,a0} - \=X_{\,b0}\,\ri,\\
&&
\=\Sigma_{\,ba1} = 2 \le\,\=X_{\,a0} -
\=X_{\,b0}\,\ri\.\=W_{\,0}\.\le\,\=X_{\,a1} - \=X_{\,b1}\,\ri
+ \le\,\=X_{\,a0} -
\=X_{\,b0}\,\ri\.\=W_{\,1}\.\le\,\=X_{\,a0} - \=X_{\,b0}\,\ri,\qquad
\end{eqnarray}
respectively, correct to the second order in $ | \, \#E_{\,HCM} \, | $~.
 Now, let us introduce the dyadic quantity
\begin{eqnarray}
&&\=\Omega = \=I +
\=\Sigma_{\,ba} \. \=D = \=\Omega_{\,0} + \=\Omega_{\,1} \,
 | \, \#E_{\,HCM} \, |^2\,,
\end{eqnarray}
such that
\begin{eqnarray}
&& \=\Omega_{\,0} = 
 \mbox{diag} \le  \Omega^x_0,
\,\Omega^y_0,\,\Omega^z_0 \ri =
 \=I +
\=\Sigma_{\,ba 0} \. \=D_{\,0}\,,\\
&&
 \=\Omega_{\,1} = 
 \mbox{diag} \le  \Omega^x_1,
\,\Omega^y_1,\,\Omega^z_1 \ri =
\=\Sigma_{\,ba 0} \. \=D_{\,1} +
\=\Sigma_{\,ba 1} \. \=D_{\,0}\,.
\end{eqnarray}
 We may then express the inverse dyadic in the form
\begin{eqnarray}
&& \=\Omega^{-1} =
\=\Omega^{-1}_{\,0} + \=\Pi \,
 | \, \#E_{\,HCM} \, |^2,
\end{eqnarray}
with nonlinear part
\begin{eqnarray}
\=\Pi &=& \frac{1}{\det \les \,
\=\Omega_{\,0}\,\ris}\,\les \, \mbox{diag} \, \Big( \,
 \Omega^y_1 \Omega^z_0 + \Omega^y_0 \Omega^z_1,\,
 \Omega^z_1 \Omega^x_0 + \Omega^z_0 \Omega^x_1,\,
 \Omega^y_1 \Omega^x_0 + \Omega^y_0 \Omega^x_1
\,\Big) - \nu \, \=\Omega^{-1}_{\,\ell\,0} \, \ris\,,
\end{eqnarray}
where
\begin{eqnarray}
\nu = \Omega^x_0 \Omega^y_0  \Omega^z_1 + \Omega^x_0 \Omega^y_1
\Omega^z_0 + \Omega^x_1 \Omega^y_0  \Omega^z_0\,.
\end{eqnarray}
Thus, the linear and nonlinear contributions of the SPFT estimate
$\=\eps_{\,ba}$ are delivered, respectively,  as
\begin{eqnarray}
&&\=\eps_{\,ba0} = \=\eps_{\,Br0} - \frac{1}{i \,\omega}\,
\=\Omega^{-1}_{\,0}\.\=\Sigma_{\,ba0}\,,\\
&&
\=\chi_{ba}=
 \=\chi_{\,Br} - \frac{1}{i \,\omega}\,\le \,
\=\Omega^{-1}_{\,0}\.\=\Sigma_{\,ba1}
+\=\Pi\.\=\Sigma_{\,ba0}\,\ri\,.
\end{eqnarray}

\section{Numerical results and discussion}

Let us  explore the HCM constitutive parameter space by means of
 a representative
numerical example: Consider the homogenization of 
a cubically nonlinear phase $a$ with linear
permittivity $\eps_{a 0} = 2 \epso$ and nonlinear susceptibility $\chi_a =
9.07571 \times
10^{-12} \epso \, \mbox{m}^2 \mbox{V}^{-2}\; 
(\equiv 6.5 \times 10^{-4} \;
\mbox{esu})$ and a linear phase $b$ with  permittivity $\eps_b
\equiv \eps_{b 0} = 12 \epso$. 
Note that the selected
nonlinear susceptibility value corresponds to that of gallium arsenide
\c{Boyd}, while selected the linear permittivity values are typical of a wide
range of insulating crystals \c{Ashcroft}.
We assume the ellipsoidal component
phase topology specified by $U_x = 1$, $U_y = 3$ and $U_z \in [0.5,15]$.
The angular frequency $\omega$  is fixed at $ 2 \pi \times 10^{10}\;
\mbox{rad s}^{-1}$ for all   calculations reported here.

The Bruggeman estimates  of the HCM relative linear 
and  nonlinear constitutive parameters are plotted in figure~1 as
functions of $f_a$ and $U_z$. The calculated constitutive parameters 
presented in figure~1
are consistent
with those calculated by Lakhtakia and Lakhtakia \c{LL01}
in a study pertaining to the Bruggeman homogenization of 
ellipsoidal inclusions with a host medium comprising spherical particles.
The linear parameters follow an approximately linear progression
between their  constraining values  at $f_a = 0$ and $f_a
= 1$. Furthermore, for the range $U_z \in [0.5,15]$, 
the linear parameters are largely (but not
completely) independent of the particulate geometry of the component
phases. This is in contrast to the nonlinear parameters which are
acutely sensitive to $U_z$.
Of special significance is the
 \emph{nonlinearity enhancement} (i.e., the 
manifestation of a higher degree of nonlinear susceptibility in the
HCM than is present in its component phases)
 which is particularly observed  at high values of
$U_z$ for $\chi^x_{Br}$ and at low values of
$U_z$ for $\chi^z_{Br}$. 
This phenomenon and its possible technological exploitation are
described elsewhere \c{Boyd96, L01,  MLW02a, MLW02b, LL01, Liao}.
In order to best consider nonlinearity enhancement, we fix 
the shape parameter $U_z = 15$ for all remaining calculations.

We turn our attention now to the bilocally-approximated SPFT calculations.
Let
\begin{equation}
\eps^{n r}_{ba 0} = \frac{\eps^n_{ba 0} - \eps^n_{Br 0 }}{\epso}, \hspace{20mm}
\chi^{n r}_{ba } = \frac{\chi^n_{ba } - \chi^n_{Br }}{\chi_a},
\hspace{20mm}
(n = x, y, z).
\end{equation}
The SPFT estimates of the HCM relative linear constitutive
 parameters $\eps^{xr,yr,zr}_{ba 0}$
and  nonlinear constitutive parameters
$\chi^{xr,yr,zr}_{ba }$
 are plotted in figures~2 and 3, respectively,  as
functions of $f_a$ and $k_0 L$. Significant differences are clear
between the Bruggeman-estimated values and the SPFT-estimated values:
The  SPFT estimates of linear constitutive parameters  provide
an additive correction to the corresponding Bruggeman parameters,
whereas for the nonlinear constitutive parameters the SPFT estimates
provide a subtractive correction to the corresponding Bruggeman parameters.
Furthermore, the magnitudes of these  differences
exhibit local maxima which occur at progressively higher values of
$f_a$ as one compares the constitutive parameter components aligned with the
$x$, $y$ and $z$ coordinate axes, respectively. This trend holds for both
the real and the imaginary parts of both the linear permittivity and the
nonlinear susceptibility parameters. However, it is less
pronounced for the nonlinear constitutive
parameters.

Coherent interactions between scattering centres enclosed
within a region of linear dimensions $L$ are accommodated in the
bilocally-approximated SPFT via the two-point covariance function
\r{cov} (see figure~1). Thus,
since neither component phase $a$ nor component  phase
$b$ is dissipative, the nonzero imaginary parts of the SPFT
constititutive parameters in figures~2 and 3
 are attributable entirely to scattering losses.
Furthermore, the  magnitudes of the imaginary
parts of the constitutive parameters are observed in figures~2 and 3
 to increase as
 $L$ increases, due to the actions of 
greater numbers of scattering centres becoming correlated.

\section{Concluding remarks}

The bilocally-approximated SPFT 
for weakly nonlinear isotropic HCMs, based on spherical 
particulate geometry, has
been recently established \c{L01, MLW02a, MLW02b}.
In the present study we further advance the theory
 through considering
anisotropic, cubically nonlinear HCMs, arising
from isotropic component phases with ellipsoidal particulate geometries. 
Significant differences between the 
bilocally-approximated SPFT (i.e., second-order theory) and the
Bruggeman homogenization formalism 
(i.e., zeroth-order theory)~---~which depend upon the
underlying particulate geometry~---~have
emerged. In particular, 
nonlinearity enhancement is  predicted to a lesser degree with the
SPFT than with the Bruggeman homogenization formalism.
The importance of taking into account the 
distributional statistics of the
HCM component phases is thereby further emphasized.

\vspace{8mm}
\noindent {\bf Acknowledgements:}
This study was partially carried out during a visit to the
Department of Engineering Science and Mechanics at Pennsylvania State
University. The author acknowledges the financial support of  \emph{The Carnegie Trust for the
Universities of Scotland} and thanks Professors Akhlesh Lakhtakia
(Pennsylvania State University) 
for suggesting the present study
and Werner S. Weiglhofer (University
of Glasgow) for numerous  discussions regarding homogenization.

\vspace{30mm}

\newpage

\begin{center} {\bf Appendix} \end{center}
 
Consider the principal value integral term $\=W$  
\r{W_def} which was expressed as
the sum $\=W_{\,0} + \=W_{\,1}\, | \, \#E_{\,HCM} \, |^2$ in \S3.
Here we develop expressions for the linear  component
$\=W_{\,0}$ and the nonlinear component $\=W_{\,1}$, appropriate to
the homogenization scenario of \S2.

Let us begin with 
 the  following straightforward
specialization of the evaluation of $\=W$  for bianisotropic
HCMs \c{MLW00, MLW01a}
\begin{eqnarray}
&& \=W = 
\frac{f_a f_b}{2 \pi^2 \, i \omega} \int d^3 \#q \;\; 
\frac{ (q/\omega)^2 \, \=\alpha + \=\beta}{(q/\omega)^4 \, t_C +
(q/\omega)^2 \, t_B + t_A}\, 
\le \, \frac{\sin  qL}{q} - L \cos q L \, \ri\,. \l{W_3d}
\end{eqnarray}
For the weakly nonlinear homogenization  outlined in \S2,
the scalar terms $t_A$, $t_B$ and $t_C$ in \r{W_3d},
along with their linear and nonlinear decompositions, 
 are given by
\begin{eqnarray}
&&  t_A = \muo^3\, \det \les \, \=\eps_{\,Br} \, \ris
=  t_{A0} + t_{A1}\, | \, \#E_{\,HCM} \, |^2,\\
&& t_B = \muo^2 \, \lec \mbox{tr} \les \, \=\eps^{adj}_{\,Br}\.\=A\,\ris - 
\le \, \mbox{tr} \les \, \=\eps^{adj}_{\,Br} \, \ris \, \mbox{tr} \les
\, \=A \, \ris \, \ri \, \ric
=  t_{B0} + t_{B1}\, | \, \#E_{\,HCM} \, |^2,\\
&& t_C = \muo \,
\mbox{tr} \les \, \=A \, \ris \,
 \mbox{tr} \les \, \=\eps_{Br}\. \=A \, \ris 
 =  t_{C0} + t_{C1}\, | \, \#E_{\,HCM} \, |^2\,,
\end{eqnarray}
wherein
\begin{eqnarray}
&& t_{A0} =  \muo^3\, \det \les \, \=\eps_{\,Br0} \,
\ris\,,\hspace{10mm}
 t_{A1} =  \muo^3\, \le \, \chi^x_{Br} \, \eps^y_{Br0}\,
\eps^z_{Br0} +
 \eps^x_{Br0} \, \chi^y_{Br}\, \eps^z_{Br0} +
 \eps^x_{Br0} \, \eps^y_{Br0}\, \chi^z_{Br}\,
 \ri\,,\quad  \\
&& t_{B0} = \muo^2 \, \lec \mbox{tr} \les \, \=\eps^{adj}_{\,Br0}\.\=A\,\ris - 
\le \, \mbox{tr} \les \, \=\eps^{adj}_{\,Br0} \, \ris \, \mbox{tr} \les
\, \=A \, \ris \, \ri \, \ric\,,\\
&& t_{B1} = \muo^2 \, \lec \mbox{tr} \les \, \=\Upsilon\.\=A\,\ris - 
\le \, \mbox{tr} \les \, \=\Upsilon \, \ris \, \mbox{tr} \les
\, \=A \, \ris \, \ri \, \ric\,,  \\
&& \=\Upsilon = \mbox{diag}\,\le \,
 \chi^y_{Br} \, \eps^z_{Br0} +
\eps^y_{Br0} \, \chi^z_{Br}, \, 
 \chi^z_{Br} \, \eps^x_{Br0} +
\eps^z_{Br0} \, \chi^x_{Br}, \, 
 \chi^x_{Br} \, \eps^y_{Br0} +
\eps^x_{Br0} \, \chi^y_{Br} \, 
\ri \,,\\
&& t_{C0} = \muo \,
\mbox{tr} \les \, \=A \, \ris \,
 \mbox{tr} \les \, \=\eps_{Br0}\. \=A \, \ris \,,\hspace{15mm}
 t_{C1} = \muo \,
\mbox{tr} \les \, \=A \, \ris \,
 \mbox{tr} \les \, \=\chi_{Br}\. \=A \, \ris \,.
\end{eqnarray}
Similarly, the dyadic quantities $\=\alpha$ and $\beta$ in \r{W_3d},
along with their linear and nonlinear decompositions, 
are given by
\begin{eqnarray}
&& \=\alpha = 
\muo^2 \,\les \,  \le 2 \,\=\eps_{\,Br} - \mbox{tr} \les \, \=\eps_{\,Br} \, \ris
\, \=I \, \ri\. \=A - \mbox{tr} \les \,\=\eps_{\,Br}\.\=A\,\ris \, \=I\,
\ris - \muo \, \frac{t_B}{t_C} \, 
\mbox{tr} \les \, \=A\, \ris \, \=A
=
\=\alpha_{\,0} + \=\alpha_{\,1}\, | \, \#E_{\,HCM} \,
|^2,\nonumber \\ && \\
&&\=\beta = 
\muo^3 \, \=\eps^{adj}_{\,Br} - \muo \, \frac{t_A}{t_C}\,\mbox{tr} \les \, \=A\, \ris \, \=A
=
\=\beta_{\,0} + \=\beta_{\,1}\, | \, \#E_{\,HCM} \, |^2,
\end{eqnarray}
with
\begin{eqnarray}
&& \=\alpha_{\,0} = 
\muo^2 \,\les \,  \le 2 \,\=\eps_{\,Br0} - \mbox{tr} \les \, \=\eps_{\,Br0} \, \ris
\, \=I \, \ri\. \=A - \mbox{tr} \les \,\=\eps_{\,Br0}\.\=A\,\ris \, \=I\,
\ris - \muo \, \frac{t_{B0}}{t_{C0}} \, 
\mbox{tr} \les \, \=A\, \ris \, \=A\,,\\
&& \=\alpha_{\,1} = 
\muo^2 \,\les \,  \le 2 \,\=\chi_{\,Br} - \mbox{tr} \les \, \=\chi_{\,Br} \, \ris
\, \=I \, \ri\. \=A - \mbox{tr} \les \,\=\chi_{\,Br}\.\=A\,\ris \, \=I\,
\ris - \muo \, \frac{t_{B1} - \le t_{B0}/t_{C0} \ri t_{C1}}{t_{C0}} \, 
\mbox{tr} \les \, \=A\, \ris \, \=A\,,\nonumber \\ && \\
&&\=\beta_{\,0} = 
\muo^3 \, \=\eps^{adj}_{\,Br0} - \muo  \frac{t_{A0}}{t_{C0}}\,\mbox{tr} \les
\, \=A\, \ris \, \=A\,,\hspace{10mm}
\=\beta_{\,1} = 
\muo^3 \, \=\Upsilon - \muo \,
 \frac{t_{A1} - \le t_{A0}/t_{C0} \ri t_{C1}}{t_{C0}} 
\,\mbox{tr} \les
\, \=A\, \ris \, \=A\,.
\end{eqnarray}

In the long-wavelength regime, i.e.,  $|\,d_\pm\,| \ll 1$ ,
the  application of residue calculus to \r{W_3d} delivers
\begin{eqnarray}
&& \=W = 
\frac{f_a f_b \, \omega}{4 \pi i}
\int^{2 \pi}_{0}  d \phi  \int^{\pi}_0
 d \theta\;\; 
\frac{\sin \theta}{3 \, \Delta}
\lec \, \frac{1}{\omega^2} \les \,
 \frac{3}{2}\, \le \, d^2_+ -
d^2_- \, \ri + i \le \, d^3_+ - d^3_- \, \ri
\ris
  \=\alpha +
i \,\le \, \frac{d^3_+}{\kappa_+} - 
\frac{d^3_-}{\kappa_-} \, \ri  \=\beta \, \ric\,,\nonumber \\ &&
\end{eqnarray}
where we have introduced
\begin{eqnarray}
&& \Delta = 
\sqrt{ t^2_B - 4 t_A t_C} = \Delta_0 + \Delta_1 \, | \, \#E_{\,HCM} \,
|^2,\\
&& \kappa_\pm = \omega^2 \, \frac{ - t_B \pm \Delta}{2\, t_C} = \kappa_{0 \pm} +
\kappa_{1 \pm}  \, | \, \#E_{\,HCM} \,
|^2,\\
&& d_\pm = L \sqrt{\kappa_\pm} = d_{0 \pm} + d_{1 \pm}\,|\, \#E_{\,HCM} \,
|^2,
\end{eqnarray}
with linear and nonlinear parts
\begin{eqnarray}
&& \Delta_0 = \sqrt{ t^2_{B0} - 4 t_{A0} t_{C0}}\,,\hspace{15mm}
 \Delta_1 =  \frac{t_{B0} t_{B1} - 2 \, \le \, t_{A1}
t_{C0} + t_{A0} t_{C1} \, \ri}{\Delta_0}\,,\\
&& \kappa_{0 \pm} = \omega^2\,\frac{ - t_{B0} \pm \Delta_0}{2\, t_{C0}}\,,
\hspace{15mm}
\kappa_{1 \pm} = \omega^2\, \frac{\le \, -t_{B1} \pm \Delta_1 \, \ri -   2 t_{C1}
\, \le \kappa_{0\pm} / \omega^2 \ri
}{2 \,  t_{C0}}\,,\\
&& d_{0 \pm} = L \sqrt{\kappa_{0 \pm}}\,, \hspace{25mm} d_{1 \pm} =
L \frac{\kappa_{1 \pm}}{2 \sqrt{\kappa_{0 \pm}}}\,.
\end{eqnarray}
The linear and nonlinear components of $\=W$ are thereby given as
\begin{eqnarray}
 \=W_{\,0} &= &
\frac{f_a f_b \, \omega}{4 \pi i}
\int^{2 \pi}_{0}  d \phi  \int^{\pi}_0
 d \theta\;\; 
\frac{\sin \theta}{3 \, \Delta_0}
\le \, \frac{\tau_{\alpha}}{\omega^2} \,
\=\alpha_{\,0} 
+ \tau_{\beta } \,
 \=\beta_{\,0} \, \ri \,, \l{W0}
\end{eqnarray}
and
\begin{eqnarray}
 \=W_{\,1} &=& 
\frac{f_a f_b \, \omega}{4 \pi i}
\int^{2 \pi}_{0}  d \phi  \int^{\pi}_0
 d \theta\;\; \sin \theta \; \times \nonumber \\ &&
\Bigg\{ 
\frac{1}{3 \, \Delta_0}
\Bigg[ \, \frac{\tau_{\alpha }}{\omega^2} \,
\=\alpha_{\,1}
+
 \tau_{\beta } \,
 \=\beta_{\,1} + 
\frac{3}{\omega^2} \Big[
 d_{0 +} d_{1+} \le 1 + i d_{0 +} \ri
-  d_{0 -} d_{1-} \le 1 + i d_{0 -} \ri \Big] \=\alpha_{\,0} 
+
\nonumber \\ && 
i \les \frac{ d^3_{0 +}}{\kappa_{0 + }}
 \le \frac{3 d_{1 +}}{d_{0 +}} - 
\frac{\kappa_{1 +}}{\kappa_{0 +}}\ri -
\frac{ d^3_{0 -}}
{\kappa_{0 - }}
 \le \frac{3 d_{1 -}}{d_{0 -}} - 
\frac{\kappa_{1 -}}{\kappa_{0 -}}\ri
\ris \=\beta_{\,0} \, \Bigg] - 
\frac{\Delta_1}{3  \Delta^2_0}
\le \, \frac{\tau_{\alpha }}{\omega^2} \,
\=\alpha_{\,0} 
+ \tau_{\beta } \,
 \=\beta_{\,0} \, \ri \, \Bigg\}\,, \quad \l{W1}
\end{eqnarray}
respectively, 
where
\begin{eqnarray}
&& \tau_\alpha = 
 \frac{3}{2}\, \le \, d^2_{0 +} -
d^2_{0 -} \, \ri + i \le \, d^3_{0 +} - d^3_{0 -} \, \ri\,, \hspace{15mm}
\tau_\beta =
i \,\le \, \frac{d^3_{0 +}}{\kappa_{0 +}} - 
\frac{d^3_{0 -}}{\kappa_{0 -}} \, \ri \,.
\end{eqnarray}

The integrals  \r{W0} and \r{W1} 
are straightforwardly evaluated by standard (e.g., Gaussian) 
 numerical methods \c{Fortran}. In the degenerate  isotropic 
case $U_x = U_y = U_z$, the integrals  \r{W0} and \r{W1} 
yield  the analytic results of \c{MLW02a}.


\newpage

\begin{figure}[!ht]
\centering \psfull \epsfig{file=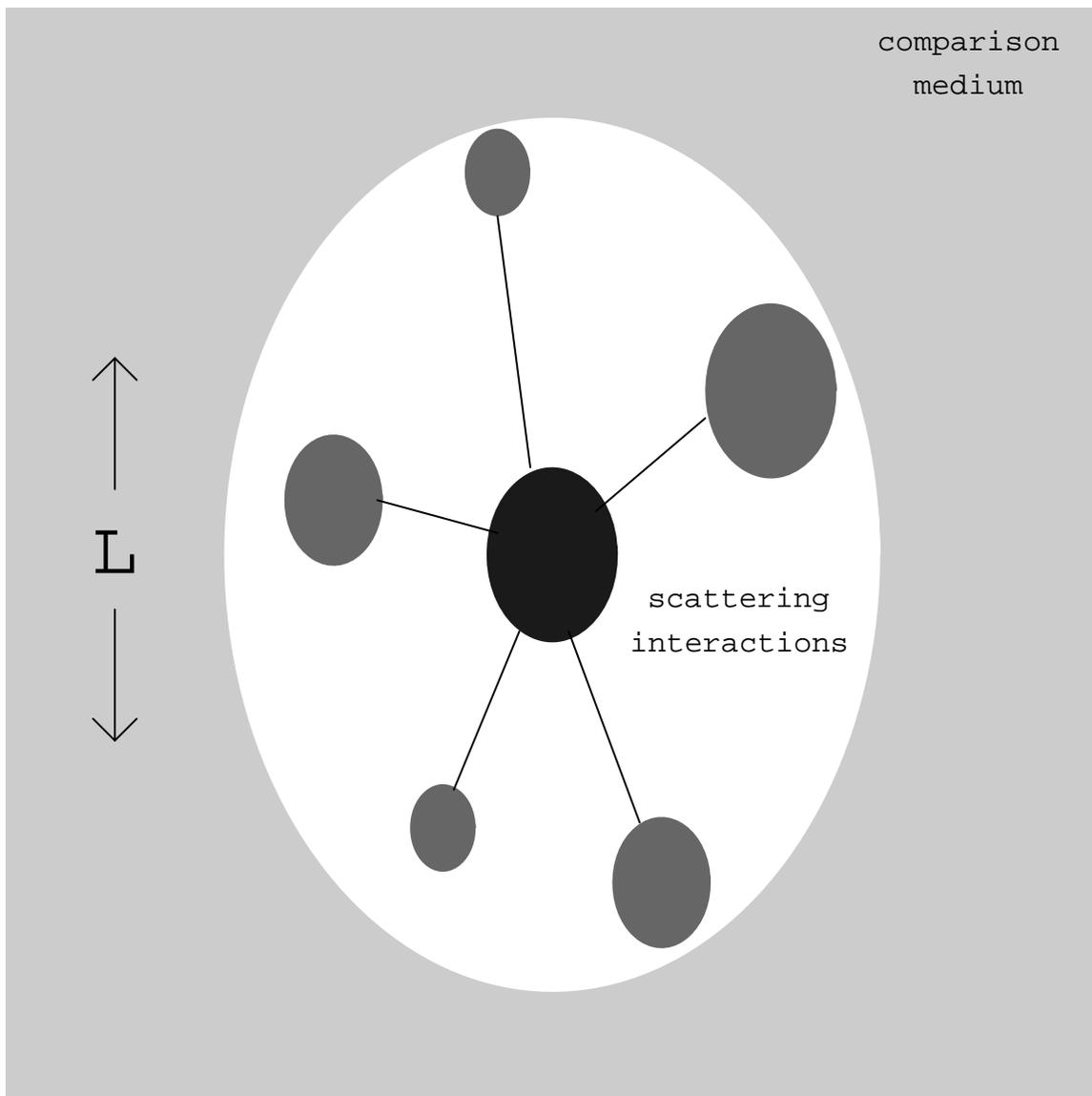,width=6.0in}
\caption{Schematic diagram  illustrating the bilocally-approximated 
SPFT for ellipsoidal component phase geometry: pair-wise 
scattering  interactions are accommodated between ellipsoidal
scattering centres contained within an ellipsoidal correlation region of
linear dimensions $L$.
}
\end{figure}

\newpage

\begin{figure}[!ht]
\centering \psfull \epsfig{file=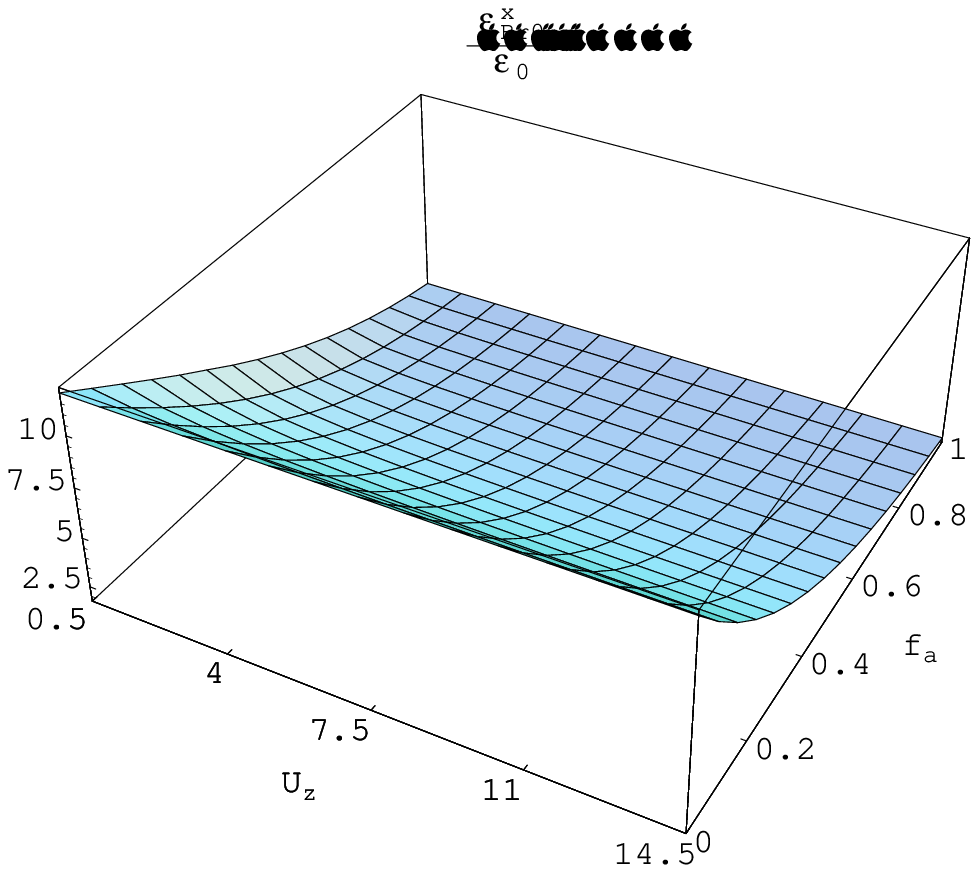,width=3.0in}
\hfill
  \epsfig{file=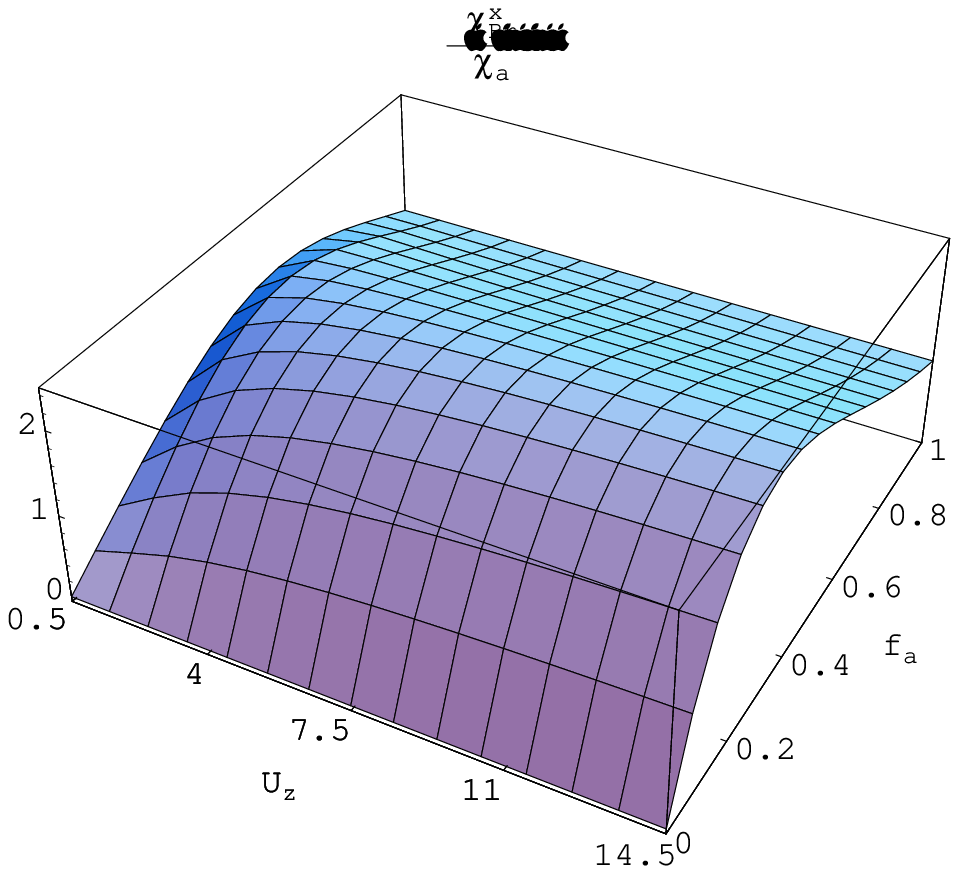,width=3.0in}
\epsfig{file=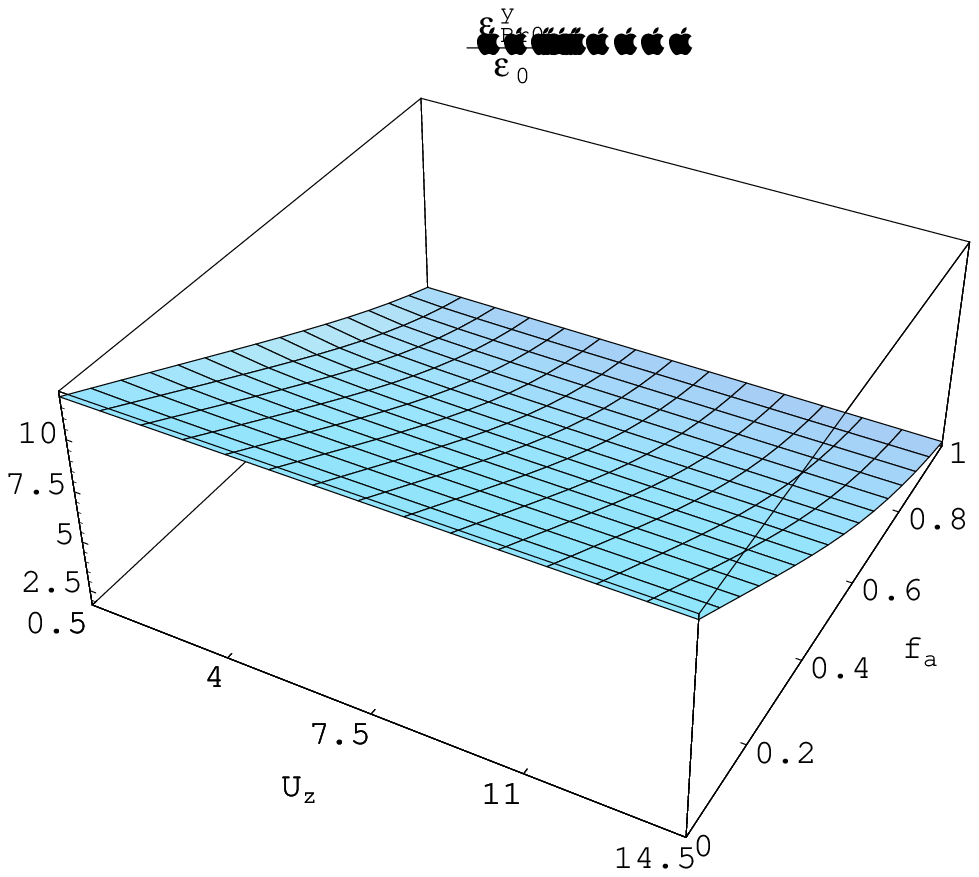,width=3.0in}
\hfill  \epsfig{file=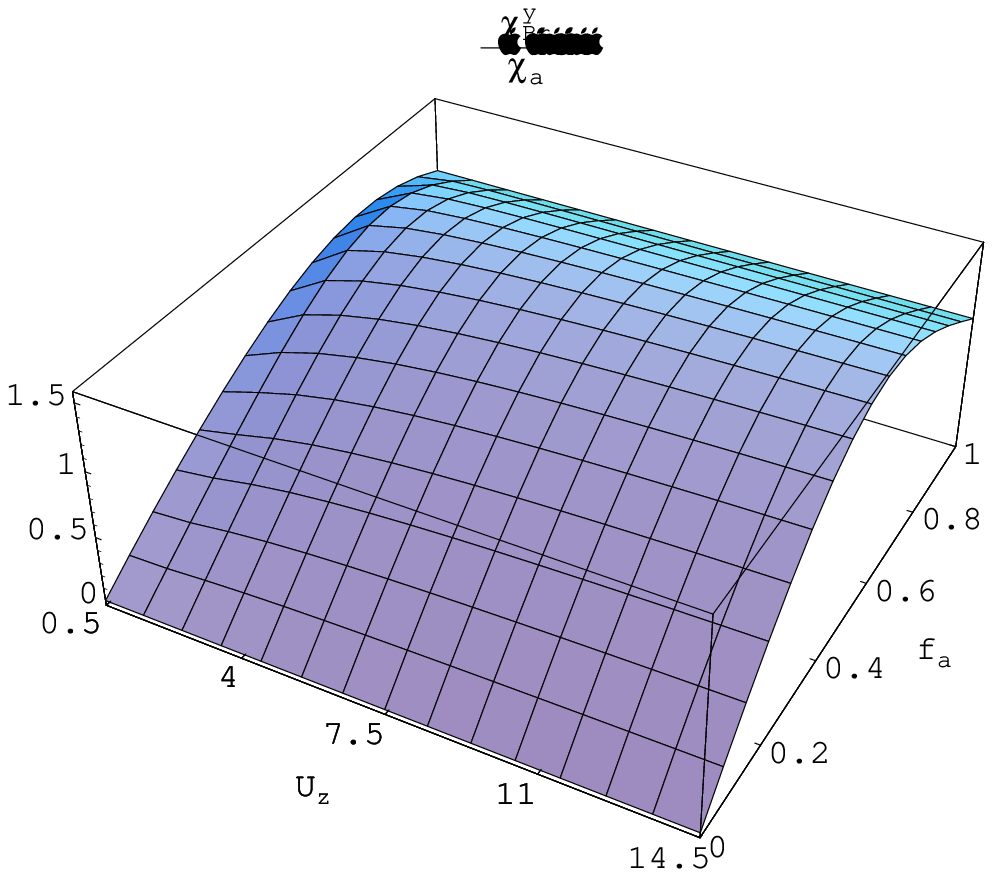,width=3.0in}
\epsfig{file=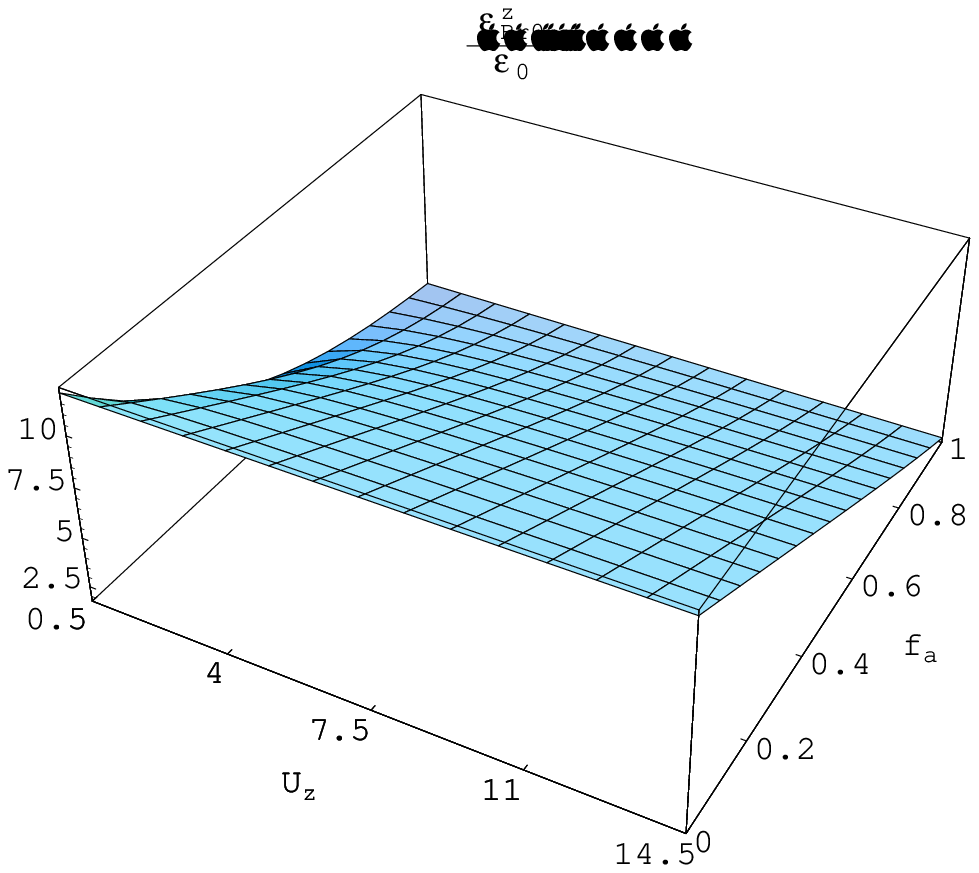,width=3.0in}
\hfill  \epsfig{file=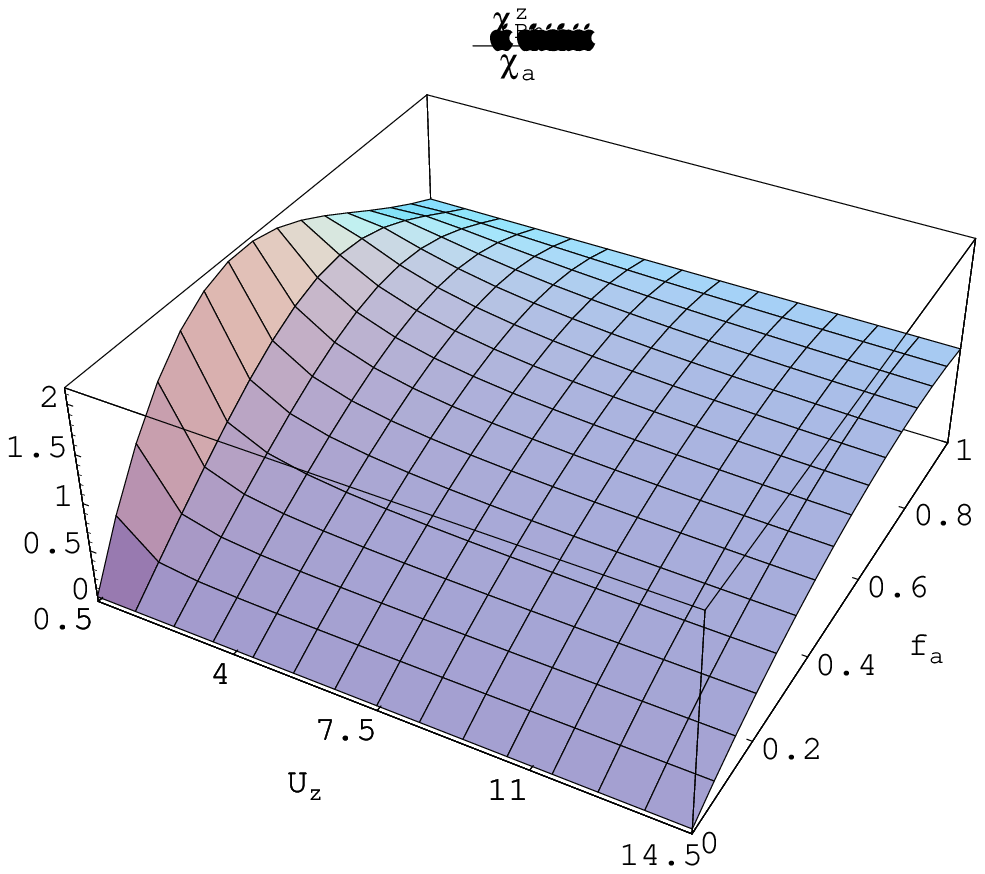,width=3.0in}
\caption{
HCM relative linear permittivity and nonlinear susceptibility parameters
calculated using the Bruggeman homogenization formalism. Component phase
parameter values: $\eps_{a 0} = 2 \epso$, $\chi_a =
9.07571 \times
10^{-12} \epso \, \mbox{m}^2 \mbox{V}^{-2}$, $\eps_b
\equiv  \eps_{b 0} = 12 \epso$, $U_x = 1$ and $U_y = 3$.}
\end{figure}

\newpage

\begin{figure}[!ht]
\centering \psfull \epsfig{file=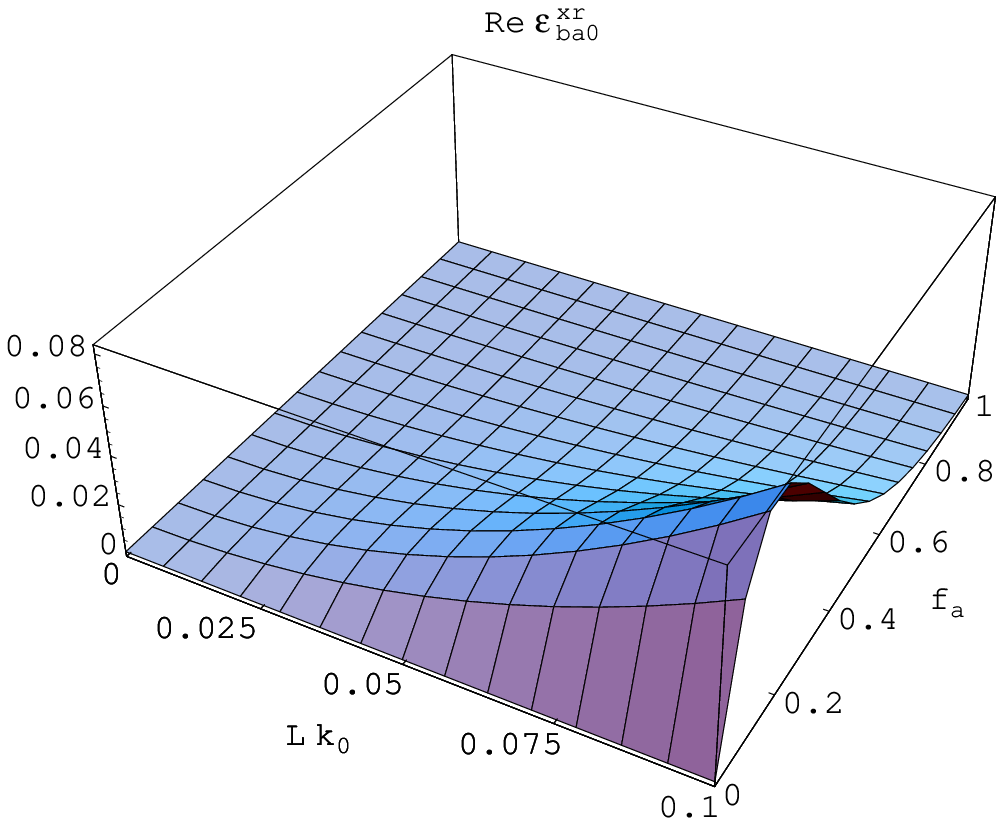,width=3.0in}
\hfill
  \epsfig{file=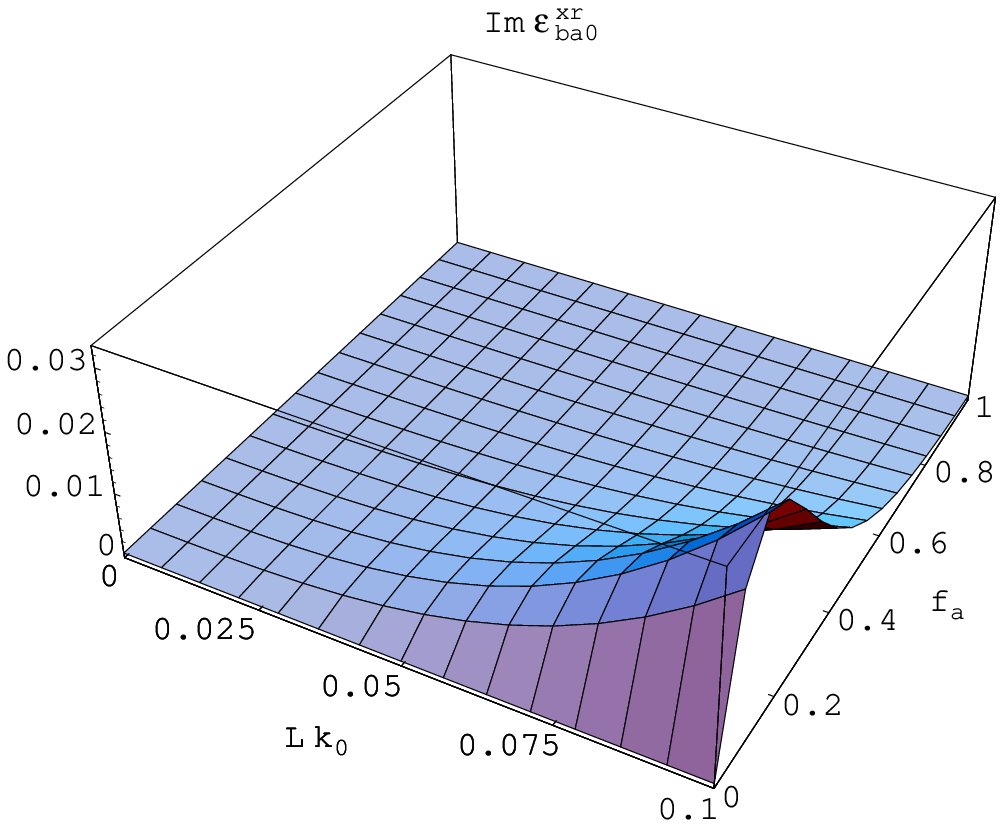,width=3.0in}
\epsfig{file=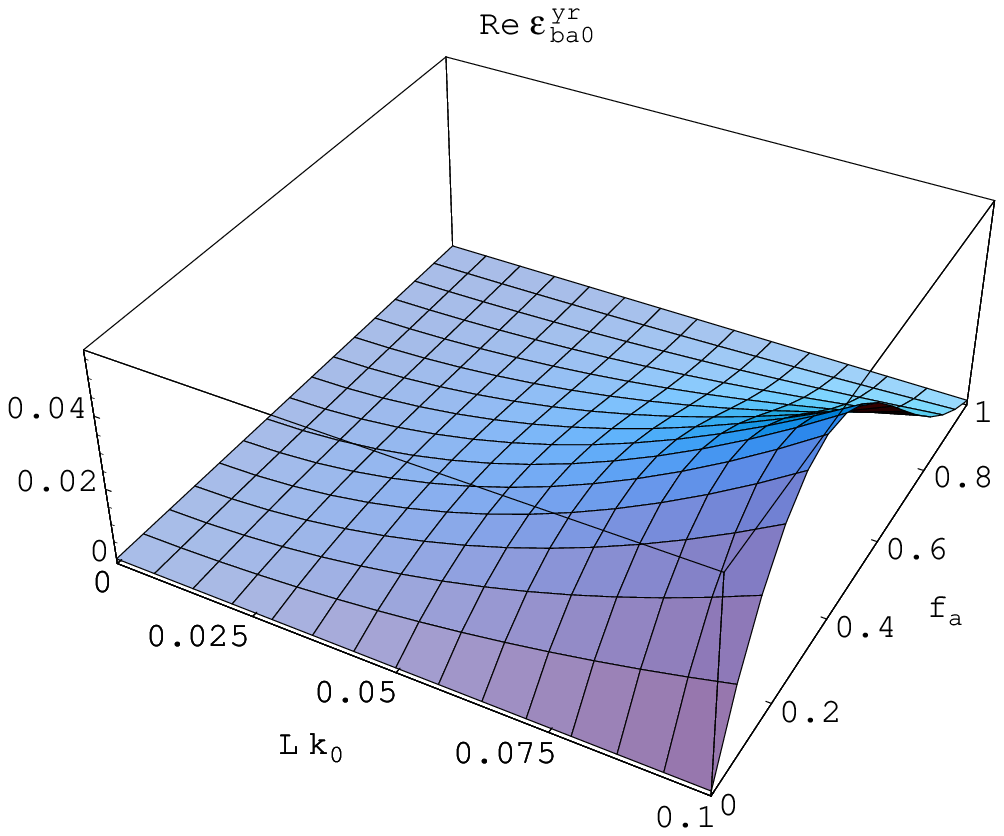,width=3.0in}
\hfill  \epsfig{file=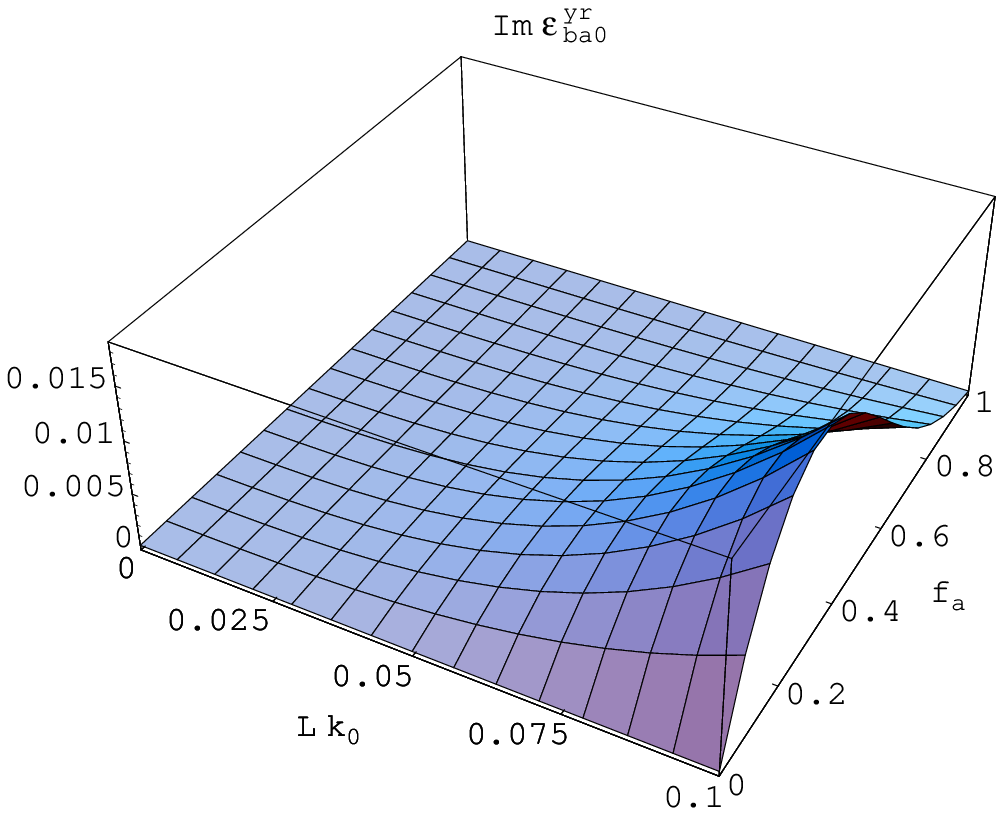,width=3.0in}
\epsfig{file=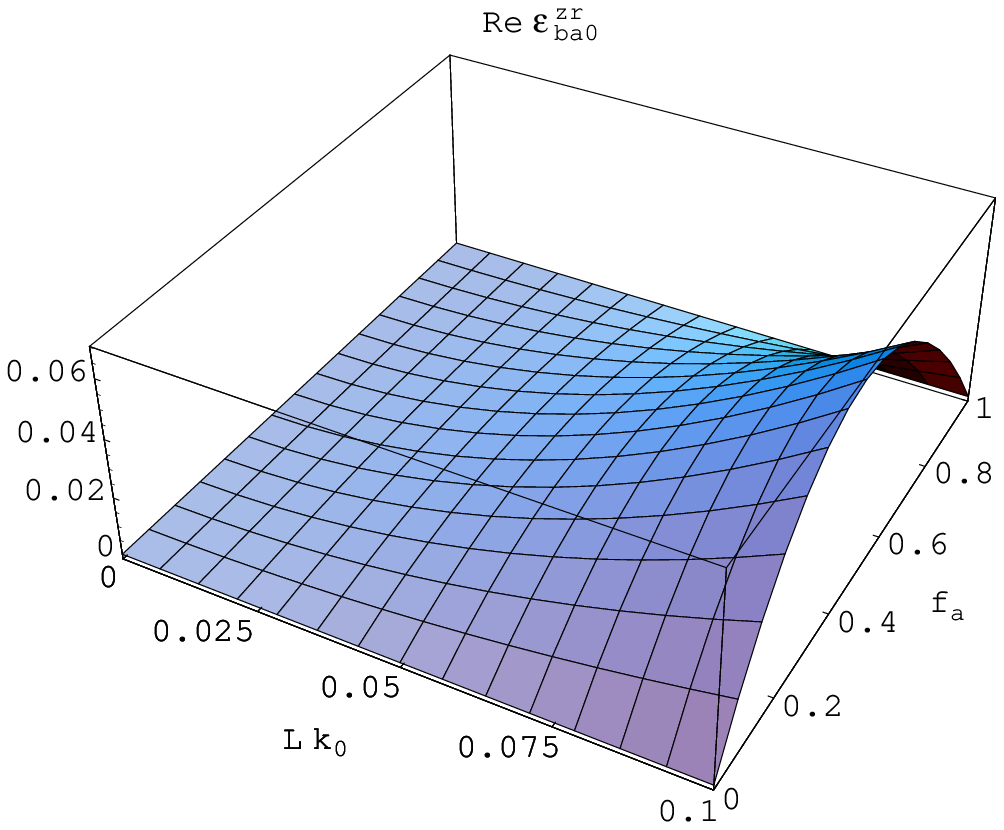,width=3.0in}
\hfill  \epsfig{file=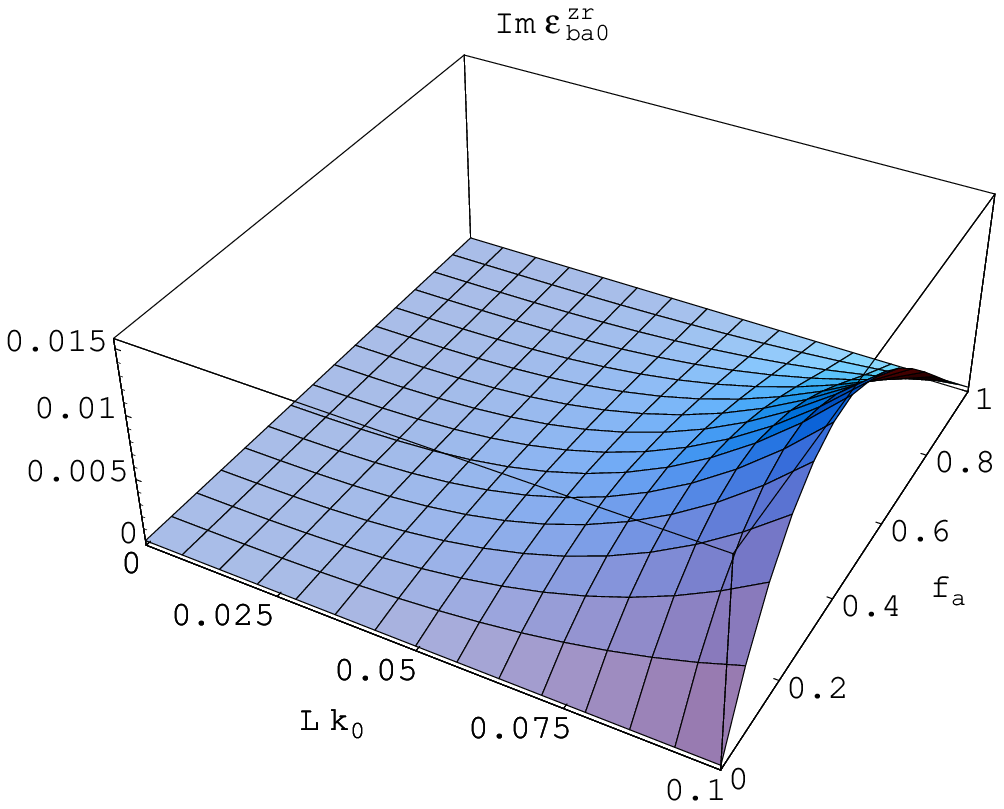,width=3.0in}
\caption{Real and imaginary parts of the 
HCM linear permittivity  parameters
calculated using the SPFT homogenization formalism. Component phase
parameter values: $\eps_{a 0} = 2 \epso$, $\chi_a =
9.07571 \times
10^{-12} \epso \, \mbox{m}^2 \mbox{V}^{-2}$, $\eps_b
\equiv  \eps_{b 0} = 12 \epso$, $U_x = 1$, $U_y = 3$ and $U_z = 15$.}
\end{figure}

\newpage

\begin{figure}[!ht]
\centering \psfull \epsfig{file=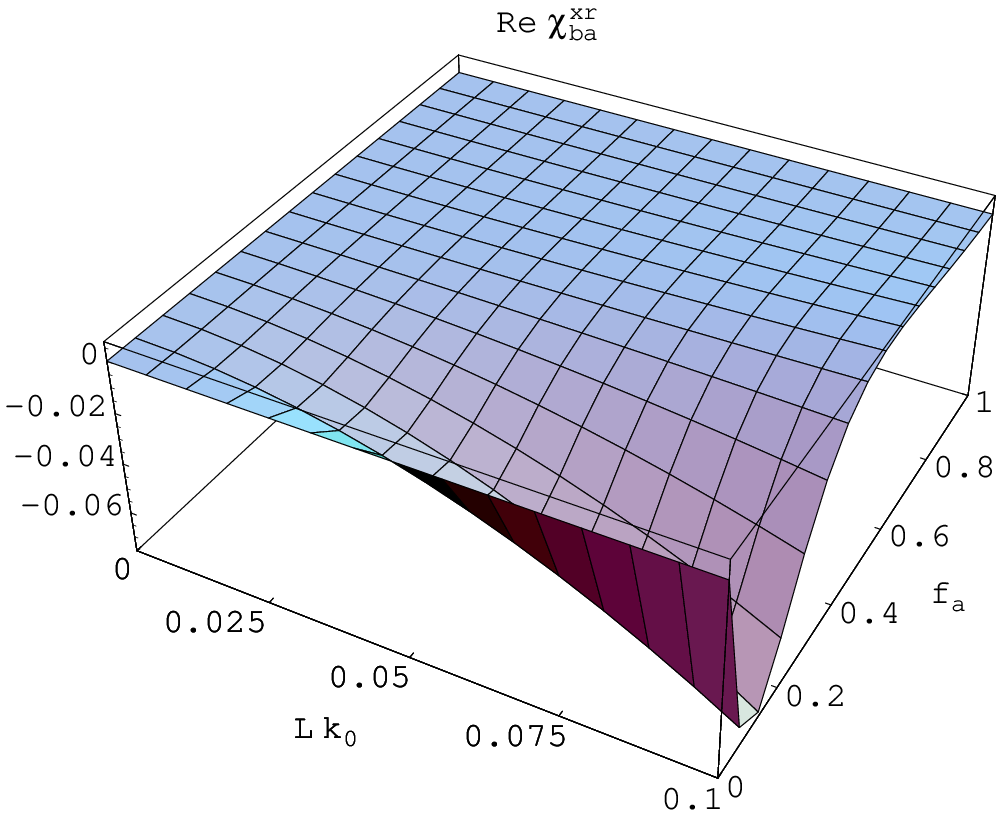,width=3.0in}
\hfill
  \epsfig{file=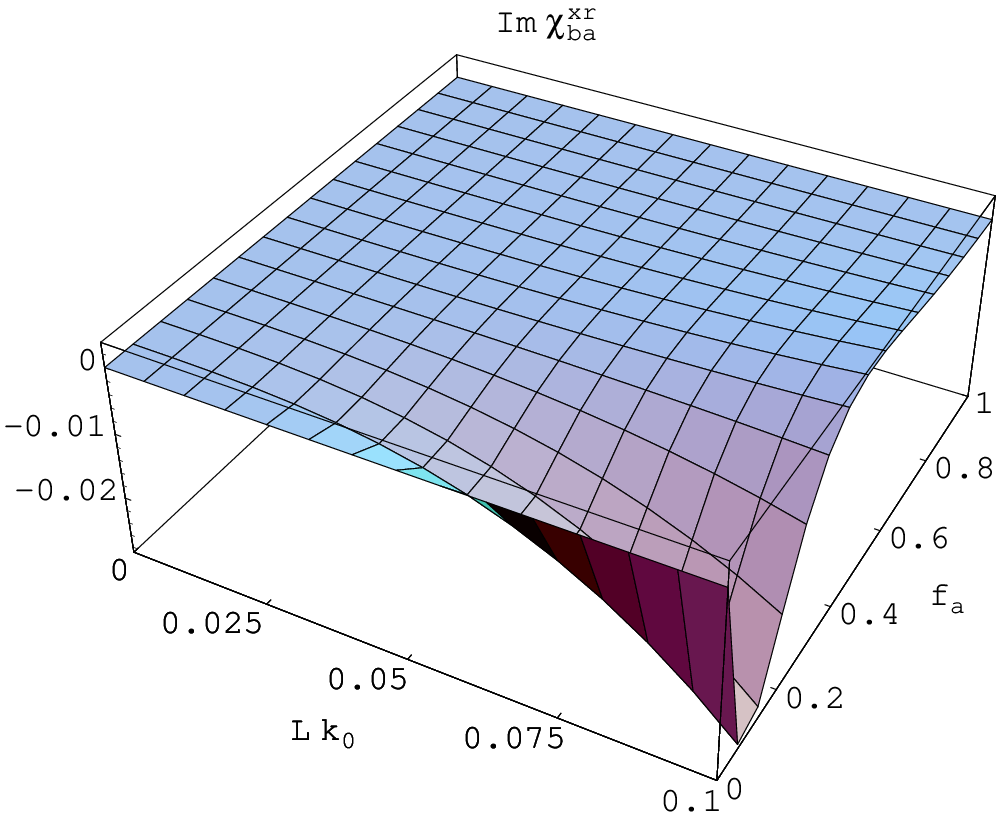,width=3.0in}
\epsfig{file=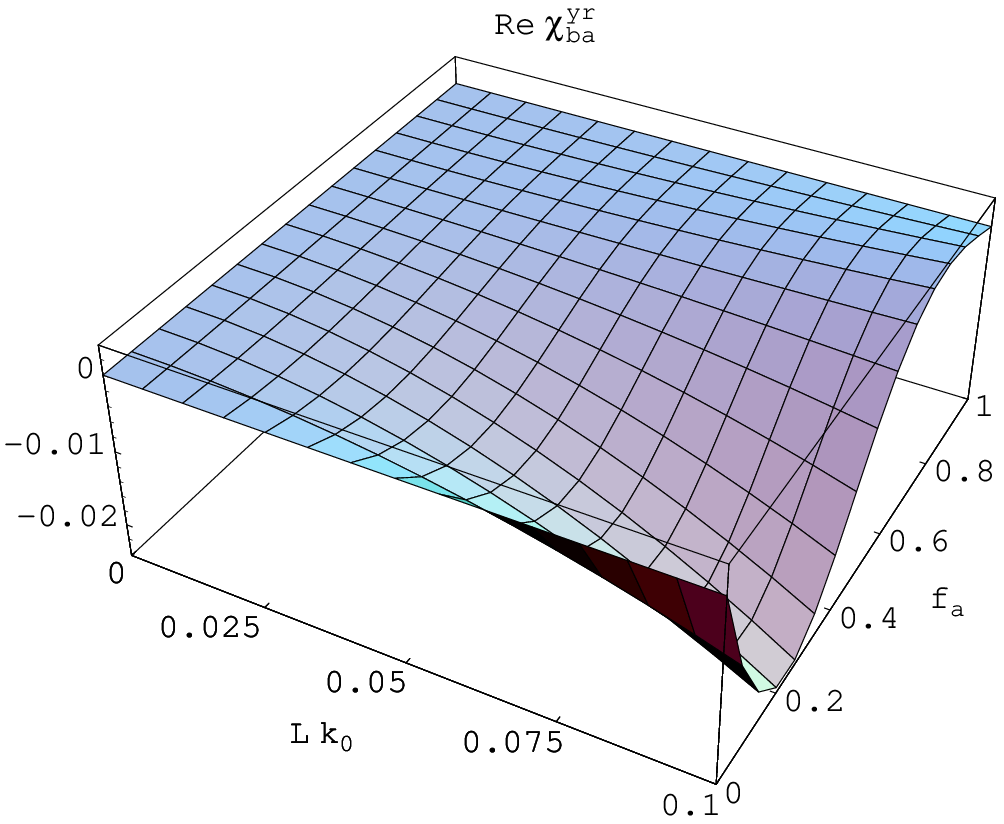,width=3.0in}
\hfill  \epsfig{file=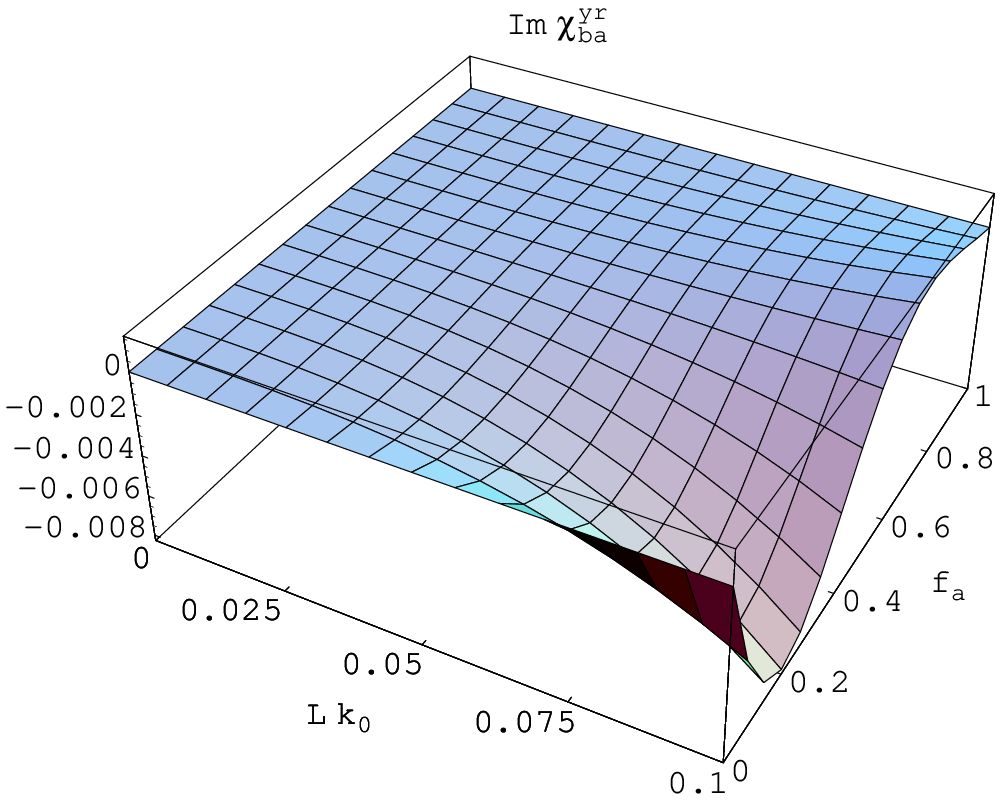,width=3.0in}
\epsfig{file=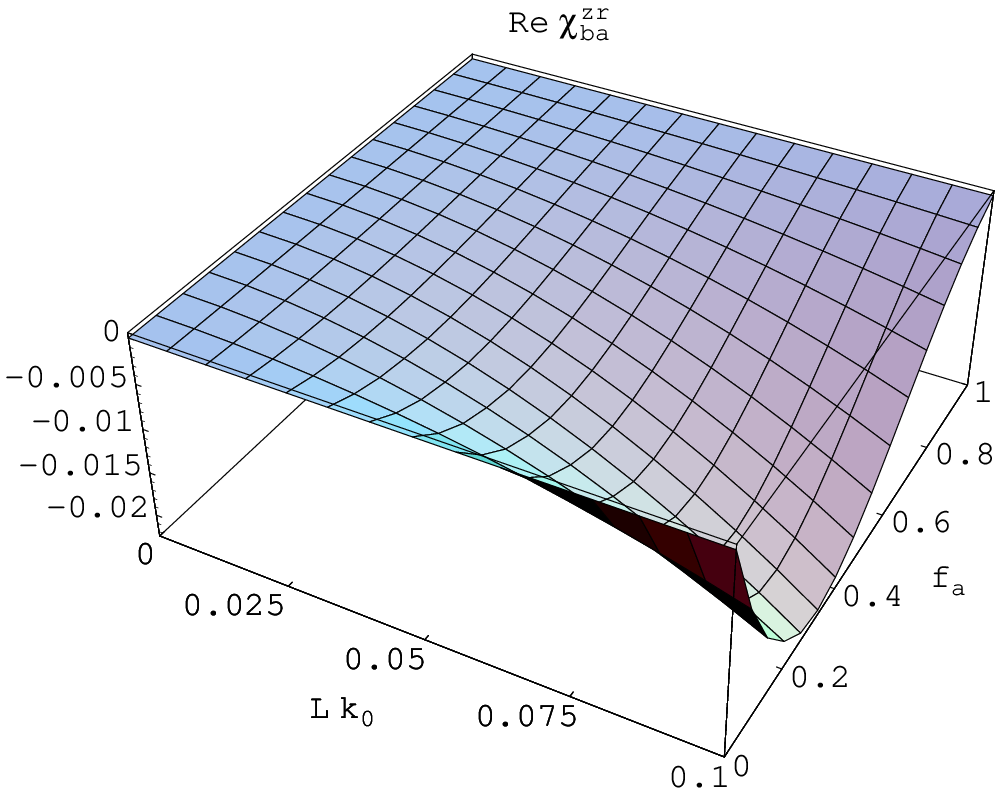,width=3.0in}
\hfill  \epsfig{file=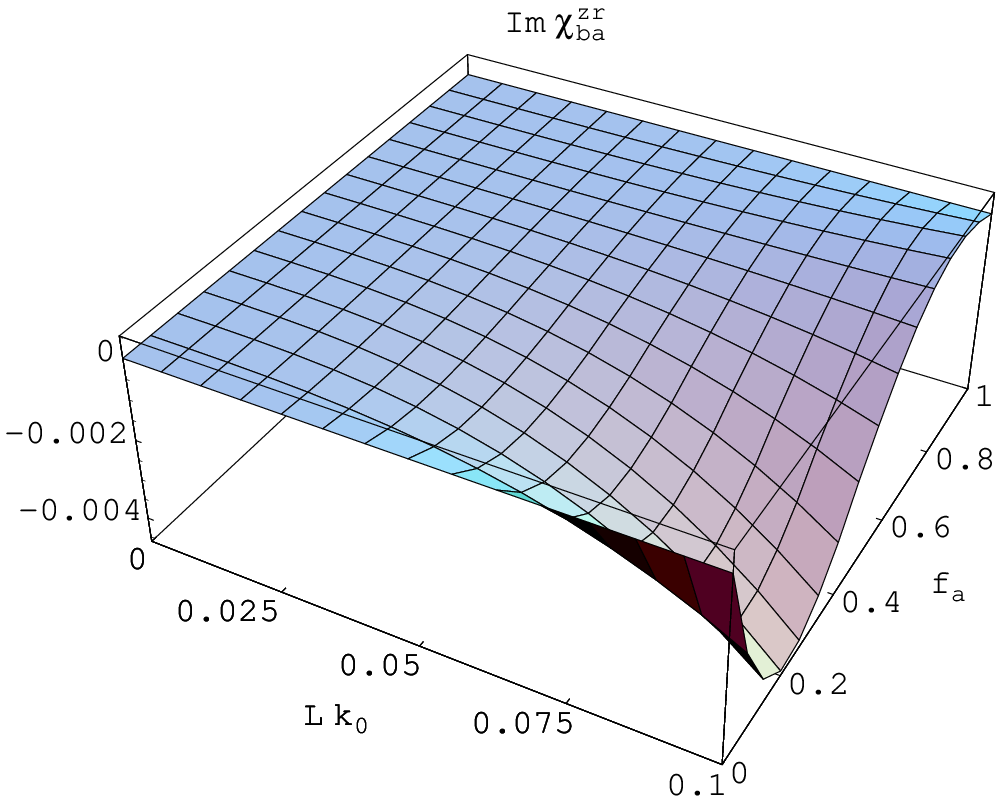,width=3.0in}
\caption{As figure~3  but  for the
HCM nonlinear susceptibility  parameters.}
\end{figure}

\end{document}